\documentclass[showpacs,prl,twocolumn,amsmath,amssymb,superscriptaddress]{revtex4-1} %-1

\usepackage{epsfig,amsmath}
\usepackage{subfigure}
\usepackage{graphicx}% Include figure files
\usepackage{dcolumn}% Align table columns on decimal point
\usepackage{stmaryrd}
\usepackage{mathrsfs}
\usepackage{pifont}
\usepackage{amsthm}
\usepackage{amssymb}
\usepackage{bm}
\usepackage{latexsym}
\usepackage{hyperref}

\newcommand{\beq}{\begin{equation}}
\newcommand{\eeq}{\end{equation}}
\newcommand{\beqa}{\begin{eqnarray}}
\newcommand{\eeqa}{\end{eqnarray}}
\begin{document}

\title{Fault-Tolerant Exact State Transmission}

\author{Zhao-Ming Wang}
%\email{Email address: mingmoon78@126.com}
\affiliation{Department of Physics, Ocean University of China, Qingdao, 266100, China}
\affiliation{Department of Theoretical Physics and History of
Science, University of the Basque Country UPV/EHU, 48008, Spain}

\author{Lian-Ao Wu}
%\email{Email address: lianao\_wu@ehu.es}
\affiliation{Department of Theoretical Physics and History of
Science, University of the Basque Country UPV/EHU, 48008, Spain}
\affiliation{IKERBASQUE, Basque Foundation for Science, 48011 Bilbao, Spain}

\author{Michele Modugno} 
\affiliation{Department of Theoretical Physics and History of
Science, University of the Basque Country UPV/EHU, 48008, Spain}
\affiliation{IKERBASQUE, Basque Foundation for Science, 48011 Bilbao, Spain}

\author{Wang Yao} 
%\email{Email address: wangyao@hkucc.hku.hk}
\affiliation{Department of Physics, The University of Hong Kong, Hong Kong, China}

\author{Bin Shao}
\affiliation{Key Laboratory of Cluster Science of Ministry of Education, and Department of Physics, Beijing Institute of Technology, Beijing 100081, China}

\date{\today}

%%%%%%%%%%%%%%%%%%%%

\begin{abstract}
We show that a category of one-dimensional XY-type models may enable high-fidelity quantum state transmissions, regardless of details of coupling configurations. This observation leads to a fault-tolerant design of a state transmission setup. 
The setup is fault-tolerant, with specified thresholds, against engineering failures of coupling configurations, fabrication imperfections or defects, 
and even time-dependent noises. We propose the implementation of the fault-tolerant scheme using hard-core bosons in one-dimensional optical lattices.  
\end{abstract}


\maketitle

{\em Introduction.---} Quantum devices require fault tolerant designs to tackle fabrication defects 
and environmental impacts, as exemplified by the fault-tolerant quantum computer. 
One of such quantum devices is a physical setup that reliably transmits a quantum state from one location to another. 
The quantum state transmission (QST) may be made through spin chains, or by ultracold atoms in 
one-dimensional (1D) optical lattices \cite{Bose2003,Wu20091,Yung2003,Wu20092,Wang2009}.
The quality of a QST depends on the coupling configurations, i. e.,
the coupling or tunneling constants $J_{i,i+1}$ as a function of site index $i$.
It has been demonstrated that a QST cannot be achieved with perfect fidelity in a uniform spin chain with XY spin couplings, 
%except for the two cases of individual couplings that are specially engineered: 
except for two cases where individual couplings are specially engineered: 
{\em perfect state transfer} (PST) \cite{Christandl2004}
and high-fidelity state transfer using weakly coupled external
qubits \cite{Wojcik05, Oh2011}. The fidelity of the latter can be achieved with arbitrary precision. 
It has been widely believed that most configurations do not enable high-fidelity QSTs \cite{Yung2003,Bayat2011}.

This letter carefully examines diverse coupling configurations. To our surprise,
the fidelity of QST is hardly determined by details
but rather on the general architecture of  $J_{i,i+1}$. Most configurations with larger values of $J_{i,i+1}$ in middle sites of the chains
work equally well.  In other words, having the particular configurations, {\em smaller on the ends and bigger in the middle},  is crucial for enabling high-fidelity state transmission, 
regardless of specific details of the couplings. We find that the fidelity is fault-tolerant against random perturbations to coupling constants and 
site energies due to the fabrication processes.
We also take into account the time-dependent random noise, which simulates the noisy effects due to environmental variables. 
Significantly, we show that the quantum states transmissions are hardly influenced by the time-dependent noise. We then specify the fault-tolerant thresholds for 
these failures, defects and noises. This observation leads to a fault-tolerant design of state transmission setup, which is
%fault-tolerant 
robust against engineering failures in the coupling configurations, fabrication imperfections or defects, and even time-dependent noise. 
Unlike the strategy in fault-tolerant quantum computer, our proposal does not require dynamical control and relies only on natures of the setup.  It is a self-protected quantum device. 

We suggest that this setup could be realized experimentally by means of ultracold bosons in 1D optical lattices, as in the hard-core limit this system can be mapped into an effective spin $\pm 1/2$ XY model \cite{lewenstein}. This system could be engineered by using the standard optical lattice technology and additional laser beams for tuning the couplings  $J_{i,i+1}$ individually. It might be more feasible to use the spatial light modulator technology, that in principle allows to create arbitrary potentials and couplings for ultracold atoms \cite{boyer,becker,henderson}, and could be used to design specific coupling configurations. The fault tolerance ensures high-fidelity QST in presence of engineering failures, fabrication imperfections or defects, and noise. 

{\em The Model and method.---} Consider a spin chain described by a Hamiltonian with couplings
 $J_{i,i+1}$ between sites $i$ and $i+1$,
 \begin{equation}
H=-\sum_{i=1}^{N-1}J_{i,i+1}(X_{i}X_{i+1}+Y_{i}Y_{i+1})+H_{\text{site}},
\label{eq:xy}
\end{equation}
where we allow $J_{i,i+1}$ to be arbitrary. The
conventional $XY$ model is a special case when $J_{i,i+1}=$ constant.
 $X _{i},Y _{i}$ are the Pauli matrices
acting on spin $i$.  $N$ is the size of the spin chain. Throughout the paper we use $N=130$ to numerically demonstrate our general results, which are almost size-independent.  We consider
an open chain of $N$ spins or two-level systems.
$H_{\text{site}}$ is the on-site potential and will be specified in our
later discussions. The {\em z-}component of the total spin or {\em magnon}
is conserved. The model is the hard-core boson limit of the Bose-Hubbard Hamiltonian
as discussed later.

In quantum information theory, quantum state transfer often refers
to transferring an unknown state, which is written as
$\left\vert \phi (0)\right\rangle $ =$\alpha \left\vert
0\right\rangle +\beta \left\vert 1\right\rangle$ and is initially in the
first spin of the chain.  Here $\left\vert 0\right\rangle
$ and $\left\vert 1\right\rangle \ $correspond to spin down and up state,
respectively. The state $\left\vert \phi (0)\right\rangle$ is
transferred to the other end as a result of free spin evolution.
The fidelity of quantum state transfer is
$f=\sqrt{\left\vert \alpha\right\vert ^{2}+\left\vert \beta
\right\vert ^{2}F}$, where $F$ is the fidelity of transmitting a {\em known}
state $\left\vert 1\right\rangle$ from the first spin to the last. When $F=1$, the transmission is exact \cite{Wu20092}. Normally, $F$
is considerably smaller than $f$. For instance, a near-perfect state transfer $f=0.97$ 
for the state with $\left\vert \beta \right\vert ^{2}=1/2$ 
only requires $F=0.9$. We will therefore focus on the fidelity, 
$F$, of {\em known} state transmissions as
discussed in \cite{Wu20092}.

The Hamiltonian can be numerically diagonalized, specifically $H=WH_{d}
W^{\dagger}$, where $H_d$ is diagonal. The propagator of the
Hamiltonian is therefore $U(t)=e^{-iHt}=We^{-iH_{d}t}W^{\dagger}$.
With this propagator, we can simulate the time evolution of
quantum states for various coupling configurations.

\begin{figure}[htbp]
\centering
\includegraphics[width=0.8\columnwidth]{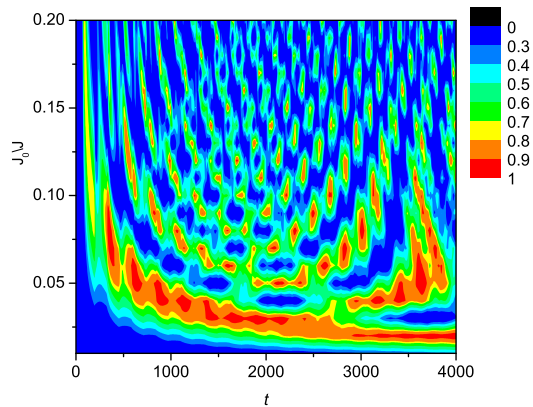}
\caption{(Color online) The contour fidelity as a function of time $t$ and $J_{0}/J$,
where $J=1.0$, $J_{1,2}=J_{N-1,N}=J_{0}$.} 
\label{fig:1}
\end{figure}

{\em Fault-tolerant QST: configurations.---} We first consider a configuration in our model, where $J_{1,2}=J_{N-1,N}=J_{0}$ and
$J_{i,i+1}=J$ elsewhere.  We term this configuration as the
weak-coupling limit and symbolize the couplings as $J^{w}(i)$.  Fig. \ref{fig:1} plots the contour
fidelity as a function of time and the ratio $J_{0}/J$. It
shows that one can reach a maximal fidelity at specific times, and that it 
decreases with $J_{0}/J$. This indicates that weaker couplings at the two ends are in favour of the
exact state transmission and quantum state transfer. For example,
$F_{\text{max}}=0.95$ when $J_{0}=0.05J$. Cautious analysis shows that the upper bound for $F_{\text{max}}>0.9$
is $J_{0}=3J/20$, which is set as a threshold for quantum state transfer ($f$ could be 0.97). The numerical treatment is consistent
with the explanations by perturbation theory as in ref. \cite{Oh2011}. On the other
hand, the time $t_{\text{MF}}$, when the first maximum
fidelity appears,  decreases with  $J_{0}/J$, which can be roughly characterized  by $
t_{\text{MF}}\propto 1/J_{0}$. 

\begin{figure}[htpb]
\centering
\includegraphics[width=0.8\columnwidth]{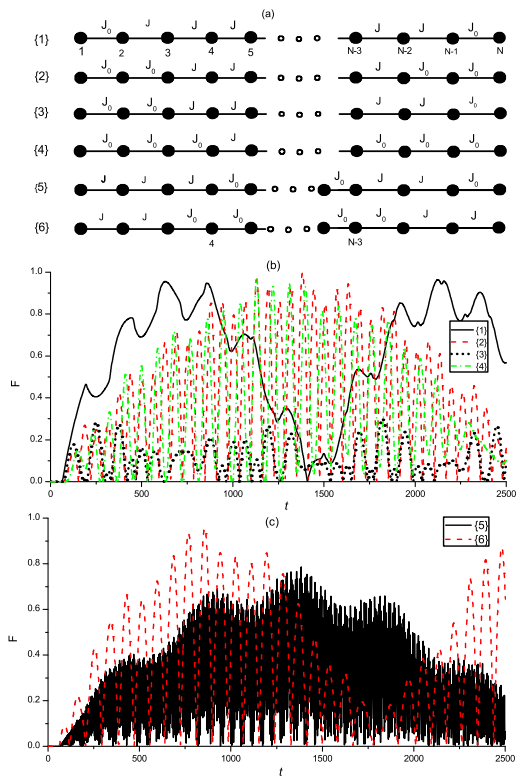}
\caption {(Color online) (a) The configurations of coupling
constants. (b) The fidelity as a function of time $t$ for cases {1-4}, where $J=1.0$, $J_{0}=0.05J$. 
(c) The fidelity as a function of time
$t$ for cases {5} and {6}, with the same parameters as in (b). } \label{fig:2a}
\end{figure}

We now study the effects of different coupling configurations in the regions of the two ends, as arranged
in Fig. \ref{fig:2a}(a), e.g., $J_{1,2}=J_{2,3}=J_{N-1,N}=J_{0}$ and  $J_{i,i+1}=J$ elsewhere in
case {3}. We compare the fidelities of the first four
configurations in Fig. \ref{fig:2a} (b), where $J_{0}/J=0.05$. Configurations {1}, {2} and {4} possesses the mirror
reflection symmetry with respect to the centre of the chain, while configuration {3} does not. The
state transmissions work equally
well for symmetric configurations. The maximum
fidelities do not vary for different configurations, though they oscillate drastically 
with increase of the numbers of $J_0$'s. On the contrary, the fidelity 
$F_{\text{max}}=0.28$ in the asymmetric configuration is very low. 

From the experimental point of view, it may be easier to implement QSTs
if sender's and receiver's sites are not at the ends but {\em inside}
the chain. Fig. \ref{fig:2a}(c) plots the fidelity vs. time when the sender is at
the fourth site and the receiver is at site $N-3$, with symmetric
configuration. Two symmetric configurations {5} and {6} in Fig. \ref{fig:2a}(c) are
compared. It shows that if both sides of the sender and receiver
attach their nearest neighbours weakly, the exact state transmission can be made. It
is interesting to note that case {5} with one $J_0$ is much worse
than case {6} with two $J_0$. The reason is that the transmission
processes to two directions (right and left) such that the strength of the transmission along one of directions is weakened.
This is verified by our numerical calculations.

{\em Fault-tolerant QST: different types of $J$.---} We now focus on
the effects of $J_{i,i+1}$ as different types of functions of sites $i$.
\begin{figure}[htpb]
\centering
\includegraphics[width=0.8\columnwidth]{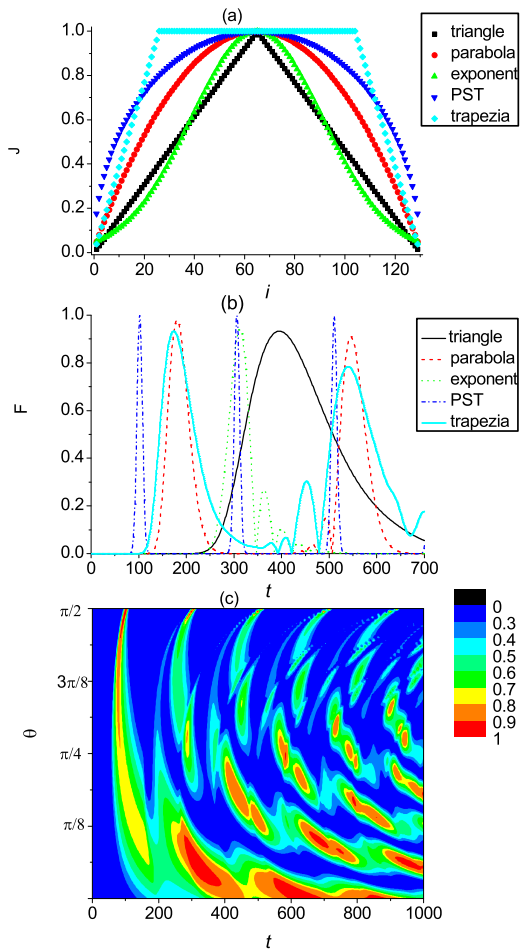}
\caption{(Color online) (a) Coupling functions $J_{i,i+1}$. (b) Corresponding fidelities as a function of time $t$ for
different $J_{i,i+1}$. (c) The contour fidelity as a function as time $t$ and $\theta$ for the interpolation $J_{i,i+1}=2\sin \theta J\sqrt{i(N-i)}/N+\cos \theta J^{w}(i)$, where $\theta$ is a parameter for the interpolation. } \label{fig:3}
\end{figure}
We have considered many types of analytical functions $J_{i,i+1}$, including
triangle $J_{i,i+1}=(2J/N)$min$(i,N-i)$; parabola $
J_{i,i+1}=-0.95J(i-N/2)^{2}/(1-N/2)^{2}+J$; exponent
$J_{i,i+1}=J\exp[\ln 0.05(i-N/2)^{2}/(1-N/2)^{2}]$; PST
$J_{i,i+1}=2\sqrt{ (i(N-i))}/N$; and trapezia $J_{i,i+1}=1.0$ (when
$i=26,27,..104$), min$(i,N-i)/26$ when $i=$ otherwise. These functions are renormalized 
such that they have the same maximal values. Fig. \ref{fig:3}(b) shows that all functions yield 
excellent fidelities, even in the worst case,
$F_{\text{max}}\approx 0.94$, with $f_{\text{max}}\approx0.985$ when
$\alpha =\beta=1/\sqrt{2}$. It is clear that an interpolation between two analytical functions will
work equally well, exemplified as follows. 

Both the weak coupling limit, $J^{w}(i)$, and the PST configurations lead to exact state transmissions with $F_{\text{max}}=1.0$.
We now study a type of interpolations between the two limits, defined by $J_{i,i+1}=2\sin \theta
J\sqrt{i(N-i)}/N+\cos \theta J^{w}(i)$. Fig. \ref{fig:3}(c) plots the time evolution of the contour fidelity for different parameter $\theta$.
$\theta=0$ and $\pi/2$ are for the two limits respectively, where exact state transmissions can be achieved. As expected, the maximum fidelity remains high for all values of $\theta$.
%\begin{figure}[htpb]
%\centering
%\includegraphics[width=2.6in]{fig5.eps}
%\caption{(Color on line) The fidelity as a function of time $t$ and
%parameter $\theta$.} \label{fig:5}
%\end{figure}

{\em Fault-tolerant QST against fabrication defects and dynamic noises.---} The random defects in fabrication is unavoidable, therefore it is crucial to
have fault-tolerant mechanism for QSTs to confront these imperfections \cite{Zwick2011}. We characterize the randomness
with $J+\gamma$rand$(i)$ for couplings $J$ and
$H_{\text{site}}=\epsilon\Sigma_{i}$ rand $(i)$  for the 
%band-broadening (BB)
on-site energies, where
$\gamma,\epsilon$ are the magnitude of the random functions in $[1,-1]$.
Our simulation shows that $J/10$ is the upper bound for $\gamma$, below which the
QST is almost perfect for both the weak coupling and PST configurations as well as interpolations.
%\begin{figure}[htpb]
%\centering
%\includegraphics[width=2.6in]{fig5.eps}
%\caption{(Color on line) (a) The fidelity as a function of time $t$
%for different
%randomness and band broadening, weak coupling distribution. $\emph{N}%
%=50$. (b)The fidelity as a function of time $t$ for different
%randomness and band broadening, PST distributions. $\emph{N}=130$}
%\label{fig:6}
%\end{figure}
%%%%%%%%%%%%%%

%{\em Fault-tolerant QST against dynamic noises}--- 
The random defects discussed above
originate from the fabrication processes. Once
formed, they will not change with time. However, it is much
more important to consider time-dependent random noises from
environments of our QST setup. Our numerical scheme
allows us to model the noise and numerically calculate its influence. We assume that the coupling constant
$J_{i,i+1}$ is perturbed by $\eta$rand$(i)$, where the random function rand$(i)$
is fixed in short time interval $\tau$. The function
rand$(i)$ is randomly different for each time interval $\tau$($=0.1$ in this paper).
In the time interval from 0 to $\tau$, the evolution operator is
$U(\tau)=\exp(-iH\tau)=W\exp(-iH_{d}\tau)W^{\dag}$.
In the next short time interval $\tau$, the Hamiltonian is changed to $H'$ and therefore $H'_d$.  Consequently, the total evolution operator in time interval
$[0,2\tau]$ is $U(2\tau)=W'\exp(-iH'_{d}\tau)W'^{\dag}W\exp(-iH_{d}\tau)W^{\dag}$. 
Continuing with the same procedure, we can numerically simulate $U(n\tau)$ with arbitrary step $n$. 

\begin{figure}[htpb]
\centering
\includegraphics[width=0.8\columnwidth]{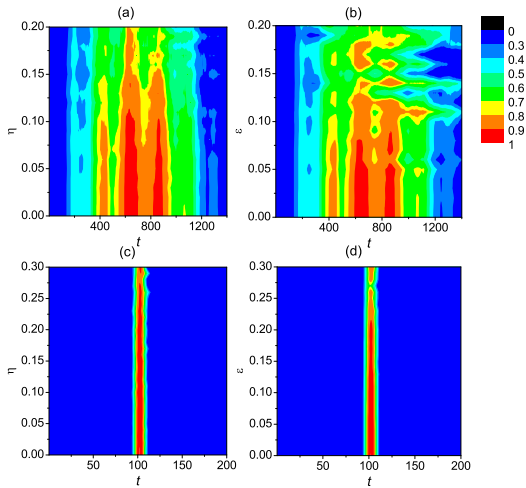} %
\caption {(Color online) The fidelity as a function of time $t$ for
time-independent noise. (a) Weak coupling limit with couplings perturbed by noise, (b) weak coupling limit with on-site noise, (c)
PST with couplings perturbed by noise, (d) PST with on-site noise.}
\label{fig:4}
\end{figure}

This numerical method helps to
exactly simulate the time-dependent random noise embedded in quantum state transfer. 
Fig. \ref{fig:4} shows the threshold with which the QST works equally well. 
The upper bound of $\gamma$ and $\epsilon$ may be $J/10$ (even bigger for the PST configuration), 
though the inhomogeneous broadening in on-site energies ruins QSTs slightly more, in particular for the weak coupling limit. It is interesting to note that the thresholds are the same for both time dependent noise
and time-independent random defects.

To understand this fault-tolerance behaviour, we can break the QST into three stages: (1) sender qubit emits its spin excitation into the spin chain in a timescale $\sim J_0^{-1}$; (2) the excitation propagates in the spin chain channel  towards the receiver qubit with propagation time determined by the spin-wave dispersion; (3) receiver qubit absorbs the spin excitation. The second stage is affected by the time-independent random defects and time-dependent noise, but they both conserve the spin excitation number in the spin chain. Thus, the spin excitation emitted by the sender can still reach the receiver side in the presence of defects or noises as long as the disorders are not strong enough to change the spin excitation from extended modes to localized modes. 
%The propagation time in the channel can be changed as the disorders can affect the dispersion relation. 
 
% INCOMPLETE CAPTION

{\em An experimental proposal. ---} Let us now consider a specific experimental implementation by means of single-species ultracold bosons in a 1D tight-binding optical lattice \cite{bloch}. An optical lattice can be created by using a retroreflected laser beam of wavelength $\lambda$, and is described by a potential $V(x)= s E_{R} \sin^{2}(kx)$, with $k=2\pi/\lambda$ and the recoil energy $E_{R}=\hbar^{2}k^{2}/2m$.
The bosons are also subjected to a transverse confinement providing the 1D geometry.
In the tight-binding regime this system is described by the following Bose-Hubbard Hamiltonian \begin{equation} 
\label{eq:bh}
\hat{H}_{HB} = -J\sum_{i=1}^{N-1}\left[\hat{a}_{i}^\dagger\hat{a}_{i+1} + h.c.\right]+ \frac{U}{2}\sum_{i}\hat{n}_{i}\left(\hat{n}_{i}-1\right),
\end{equation}
with a uniform tunneling rate, $J/E_{R}= (4/\sqrt{\pi})s^{0.75}\exp(-2.07\sqrt{s})$ \cite{gerbier}. In the hard boson limit, $U\gg J$,  one can consider states with at most one boson per site and use the presence/absence of a boson at a site to encode a quibit. The last term in (\ref{eq:bh}) vanishes.  In this regime the Hamiltonian can be cast  in the form of the XY model in (\ref{eq:xy}) \cite{lewenstein}. Next one may tune the couplings $J$ individually by focusing additional laser beams perpendicular to the lattice direction in correspondence of single sites. For sufficiently high intensities it is possible to create box-like barriers \cite{raizen}. While this technique can create uniform couplings $J$ perfectly, the precise control of individual couplings, such as lattice ends or sites with a different coupling constant $J_{0}$, may be experimentally challenging. The major fault could happen when one uses the transverse lasers to address specific sites and locks these lasers in position, which may be tolerated in our scheme. On the other hand, a more promising technique is provided by the spatial light modulator (SLM) technology, that in principle allows to design arbitrary potentials for ultracold atoms \cite{boyer,becker,henderson}, and could be used to create one of our coupling configurations: {\em smaller on the ends and bigger in the middle}. This is a technique that is widely used for biological applications (see e. g. \cite{grier} and references therein) and that is becoming to be available in experiments with ultra cold atoms. In addition, both techniques are subject to noises, e. g., from background gas, which could be tolerated in our design.

{\em Conclusions.---} We have demonstrated high-fidelity QST for 
a variety of coupling configurations. Besides the two perfect state transfer schemes, the PST and the weak coupling, most symmetric configurations with larger values of $J_{i,i+1}$ in middle sites of the chains works equally well. We have found thresholds for enabling high-fidelity QSTs with manufacturing imperfections and even in noise channels. We have also proposed a specific experimental implementation with hard-core boson in 1D optical lattices,
designed by means of the current optical lattice technology or by using spatial light modulators. 
%This setup is expected to be robust against engineering failures in the coupling configurations, fabrication defects, and even time-dependent noise.

\begin{acknowledgements}
MM thanks C. Fort and G. Modugno for useful discussions and suggestions.
This material is based upon work supported by NSFC (Grant Nos.
11005099), Fundamental Research Funds for the Central
Universities (Grant No. 201013037), an Ikerbasque Foundation Startup,
the Basque Government (grant IT472-10) and the Spanish MICINN
(Project No. FIS2009-12773-C02-02), the NSF PHY-0925174,
DOD/AF/AFOSR No.~FA9550-12-1-0001, the UPV/EHU under program UFI 11/55, and the Research Grant Council of Hong Kong.

\end{acknowledgements}

\end{document}